\documentclass[showpacs,prl,onecolumn,aps,superscriptaddress,preprintnumbers,nofootinbib]{revtex4}
\usepackage{amsmath,amssymb}
\usepackage{epsfig}
\usepackage{graphicx}
\usepackage{amsmath}
\usepackage{amsfonts}
\def\slashchar#1{\setbox0=\hbox{$#1$}     		% set a box for #1
   \dimen0=\wd0                                 	% and get its size
   \setbox1=\hbox{/} \dimen1=\wd1               	% get size of /
   \ifdim\dimen0>\dimen1                        	% #1 is bigger
      \rlap{\hbox to \dimen0{\hfil/\hfil}}      	% so center / in box
      #1                                        	% and print #1
   \else                                        	% / is bigger
      \rlap{\hbox to \dimen1{\hfil$#1$\hfil}}   	% so center #1
      /                                         	% and print /
   \fi}

\renewcommand{\vec}{\boldsymbol}
\newcommand{\beq}{\begin{equation}}
\newcommand{\eeq}{\end{equation}}
\newcommand{\bea}{\begin{eqnarray}}
\newcommand{\eea}{\end{eqnarray}}
\newcommand{\ba}{\begin{array}}
\newcommand{\ea}{\end{array}}

\def\eq#1{{Eq.~(\ref{#1})}}
\def\fig#1{{Fig.~\ref{#1}}}
\newcommand{\bas}{\bar{\alpha}_S}

\newcommand{\nn}{\nonumber}

\newcommand{\h}{\frac{1}{2}}

\newcommand{\ha}{{\cal H}}

\newcommand{\Lb}{\left(}
\newcommand{\Rb}{\right)}

\begin{document}

\title{ A new parton model for the soft interactions at high energies: the
 Odderon.	}
\author{E. ~Gotsman}
\email{gotsman@post.tau.ac.il}
\affiliation{Department of Particle Physics, School of Physics and Astronomy,
Raymond and Beverly Sackler
 Faculty of Exact Science, Tel Aviv University, Tel Aviv, 69978, Israel}
 \author{ E.~ Levin}
\email{leving@tauex.tau.ac.il, eugeny.levin@usm.cl}
\affiliation{Department of Particle Physics, School of Physics and Astronomy,
Raymond and Beverly Sackler
 Faculty of Exact Science, Tel Aviv University, Tel Aviv, 69978, Israel}
 \affiliation{Departemento de F\'isica, Universidad T\'ecnica Federico
 Santa Mar\'ia, and Centro Cient\'ifico-\\
Tecnol\'ogico de Valpara\'iso, Avda. Espana 1680, Casilla 110-V,
 Valpara\'iso, Chile} 
 \author{  I.~ Potashnikova}
\email{irina.potashnikova@usm.cl}
\affiliation{Departemento de F\'isica, Universidad T\'ecnica Federico
 Santa Mar\'ia, and Centro Cient\'ifico-\\
Tecnol\'ogico de Valpara\'iso, Avda. Espana 1680, Casilla 110-V,
 Valpara\'iso, Chile}
\date{\today}


\begin{abstract}

In this paper we discuss  the Odderon contribution in 
our
 model\cite{GLPPM} that  gives a fairly good description of 
 $\sigma_{\rm tot}$,$\sigma_{\rm el}$
and $B_{\rm el}$  for  proton-proton scattering.  We show that the 
shadowing 
corrections are large,
 and induce considerable dependence on energy for the
 Odderon contribution,  which  in perturbative QCD  does not depend 
on energy.
  This energy dependence is in agreement with
   the experimental data for $\rho = {\rm Re}/{\rm Im}$, 
 assuming that the Odderon gives a contribution of about 1 mb at
 W =  7 TeV. This  differs greatly with our 
estimates for a
 CGC based model.
 
 We reproduce the main features of the $ t$ -dependence that are measured
 experimentally: the slope of the elastic cross section at small $t$, the
 existence of the minima in $t$-dependence which is located at
 $|t|_{min}$ = 0.52 GeV 2 at W= 7 TeV; and the behaviour of the 
elastic cross 
section at $  |t| > |t|_{min}$.
The real part   of the elastic  amplitude turns out to be much 
smaller than the experimental one.
 Consequently, to achieve a description of the data, it is necessary
 to add an odderon contribution.
 \end{abstract}

\maketitle

\vspace{-0.5cm}
\tableofcontents

%\flushbottom

 \section{ Introduction}
 
 The new data of TOTEM collaboration\cite{TOTEMRHO1,TOTEMRHO2,TOTEMRHO3,
TOTEMRHO4}
drew attention to  the state with negative signature and with an
 intercept which is close to unity
 (see Refs.\cite{KMRO,BJRS,TT,MN,BLM,SS,KMRO1,KMRO2,GLP,CNPSS}).
 This state is known as  the Odderon, and it appears  naturally  in
 perturbative QCD (see  Ref.\cite{KOLEB} for the review)
 with the intercept $\alpha_{\rm Odd}\Lb t=0\Rb\,\,=\,\,1$\cite{BLV,KS}.
 In a  number of papers it was  shown that such  a state could be
  helpful for describing the experimental
 data\cite{KMRO,BJRS,TT,MN,BLM,SS,KMRO1,KMRO2,GLP,CNPSS}.

 However, in perturbative QCD   the dependence of the Odderon on  energy 
 is crucially affected by the shadowing corrections, which  lead to a
 substantial decrease of the Odderon contribution 
 with increasing energy\cite{KS,KOLEB,CLMS}. 
 
 In this paper we wish to study the shadowing corrections to the
 Odderon contribution using  the model  that we  proposed in
 Ref.\cite{GLPPM}. The model  is based  (i) on  
 Pomeron calculus in 1+1 space-time, suggested in Ref. \cite{KLL}, 
 and 
(ii) on
 simple assumptions of hadron structure, related to the impact 
parameter
 dependence of the scattering amplitude. This parton model stems from QCD,
 assuming that the unknown non-perturbative corrections lead to 
determining  
 the
 size of the interacting dipoles. The advantage of this approach is that 
it
 satisfies both the $t$-channel and  $s$-channel unitarity, and can be 
used
 for summing all diagrams of the Pomeron interaction, including  Pomeron
 loops. In other words, we can use this approach for all possible
 reactions: dilute-dilute (hadron-hadron), dilute-dense (hadron-nucleus)
 and dense-dense (nucleus-nucleus), parton systems scattering.

The model gives  a  fairly  good description of 
three 
experimental
 observables: $\sigma_{\rm tot}$,$\sigma_{\rm el}$ and  $B_{\rm el}$   for proton-proton scattering,  
 in  the eikonal model for
 the structure of hadrons at high energy.  
 The goal of this paper is  study the influence of the shadowing
 corrections on the Odderon contribution in our model.

%%%%%%%%%%%%%%%%%%%%%%%%%%%%%%%%%%%%%%%%%%%%%%%%%
\section{ The model (brief review)}  
 Our model includes three essential ingredients: (i) the new parton
 model for the dipole-dipole scattering amplitude that has been discussed
 above; (ii) the simplified one channel model that  enables  us to 
take 
 into account  diffractive production in  the low mass region, and (iii)
 the assumptions for impact parameter dependence of the initial 
conditions.

\subsection{New parton model.}
 %%%%%%%%%%%%%%%%%%%%%%%%%%%%%%%%%%%%%%%%%%%%%%%%%
The model that we  employ  \cite{GLPPM,KLL} is  based on 
three
 ingredients:
 
 1. The Colour Glass Condensate
(GCC) approach (see Ref.\cite{KOLEB} for a review), which can be
 re-written in an equivalent form as the interaction of   BFKL
 Pomerons\cite{AKLL} in a limited range of rapidities
 ( $Y \leq Y_{\rm max}$):

 \beq \label{RAPRA}
Y \,\leq\,\frac{2}{\Delta_{\mbox{\tiny BFKL}}}\,\ln\Lb
 \frac{1}{\Delta^2_{\mbox
{\tiny BFKL}}}\Rb
\eeq 
 $\Delta_{\mbox{\tiny BFKL}}$ denotes the intercept of the BFKL 
 Pomeron\cite{BFKL}. In our model $ \Delta_{\mbox{\tiny BFKL}}\,
\approx\,0.2 - 0.25$    leading to $Y_{max} = 20 - 30$, which covers
 all collider energies.     

2. The   following Hamiltonian:
  \begin{equation}\label{HNPM}
\ha_{\rm NPM}=-\frac{1}{\gamma}\bar PP\eeq
where NPM stands for ``new parton model''. $P$ and $\bar P$  denote
the BFKL
 Pomeron fields.
The fact that it is self dual is evident.  This Hamiltonian in the limit
 of small $\bar P$ reproduces the  Balitsky-Kovchegov Hamiltonian
 $\ha_{\rm BK}$
( see Ref.\cite{KLL} for details). This condition is 
 important for  determining the form of
$\ha_{\rm NPM}$.  $\gamma$ in \eq{HNPM} denotes the dipole-dipole
 scattering amplitude, which in QCD is proportional to $\bas^2$.
 
 3. The new commutation relations:
\beq\label{CRCOR}
\Big(1\,\,-\,\,P\Big)\Big(1\,\,-\,\,\bar P \Big)\,\,=\,\,(1-\gamma)\Big(1\,\,-\,\,\bar P\Big) \Big(1\,\,-\,\,P\Big)
\eeq
For small $\gamma$  and in the regime where  $P$ and $\bar P$ are also 
small, we obtain
\beq
[P,\bar P]=-\gamma +...
\eeq
consistent with the standard  BFKL Pomeron calculus (see Ref.\cite{KLL}
 for details) . 

 In Ref.\cite{KLL}, it was shown that the scattering matrix 
for the model 
is
  given by 
\begin{eqnarray}\label{classs}
S^{\rm NPM}_{m\bar n}(Y)&=&e^{\frac{1}{\gamma} \int_0^Yd\eta\left[
 \ln(1-p)\frac{\partial}{\partial \eta}\ln (1-\bar p) 
+\bar pp\right]}[1-p(Y)]^m[1-\bar p(0)]^{\bar n}|_{p(0)=1-e^{-\gamma
 \bar n};\  \bar p(Y)=1-e^{-\gamma m}}\nonumber\\
&=&[1-p(Y)]^m\,e^{\frac{1}{\gamma}\int_0^Yd\eta \left[\ln(1-\bar p)+\bar
 p\right]p}
\end{eqnarray}
where $p(\eta)$ and $\bar p(\eta)$ are solutions of the classical equations
 of motion and have the form:

 \beq \label{H03}
 P (\eta)\,=\,\frac{ \alpha +\beta e^{ (1 - \alpha) \eta} }{1 + \beta e^{ (1
 - \alpha)  \eta}}; \ \ \ \ \bar P(\eta)=   \frac{ \alpha (1+\beta e^{
 (1 - \alpha) \eta}) }{\alpha +  \beta e^{ (1 - \alpha)  \eta}};
 \eeq
 where the parameters $\beta$ and $\alpha$ should be  determined from 
the
 boundary conditions:
 \beq \label{H0BC}
 P (\eta= 0)\,=\,p_0;\,\,\,\,\,\,\,\, \bar P (\eta= Y)\,=\,\frac{\alpha}{P
 (\eta= Y)}\,=\,\bar p_0
 \eeq

%%%%%%%%%%%%%%%%%%%%%%%%%%%%%%%%%%%%%%%%%%%%%%%%
 \subsection{ Eikonal approximation } 
 %%%%%%%%%%%%%%%%%%%%%%%%%%%%%%%%%%%%%%%%%%%%%%%%%%
 In the eikonal approximation we neglect the contribution of the
 diffractive production and assume that the hadron wave function
 diagonalize the matrix of interaction. In this model
 the unitarity constraints take  the form
\beq \label{UNIT}
2\,\mbox{Im}\,A\left(s,b\right)=|A\left(s,b\right)|^2
+G^{in}(s,b),
\eeq
where $G^{in}$ denotes the contribution of all  inelastic processes.
 In \eq{UNIT} 
$\sqrt{s}=W$ denotes the energy of the colliding hadrons and $b$ denotes 
the 
impact  parameter. In our approach we used the solution to  \eq{UNIT}
given by  \eq{classs} and 
\beq \label{EIK}
A \,=\,1 - S^{\rm NPM}(Y, b)\,\,\,\equiv\,\,\,1\,\,-\,\,\exp\Big(- \Omega\Lb Y,b\Rb\Big)
\eeq

%%%%%%%%%%%%%%%%%%%%%%%%%%%%%%%%%%%%%%%%%%%%%%%%%%%
  \subsection{ The general formulae.} 
 %%%%%%%%%%%%%%%%%%%%%%%%%%%%%%%%%%%%%%%%%%%%%%%%%%% 

 {\it  Initial conditions:}
 Following Ref.\cite{GLPPM} we chose the initial conditions in the form:
\begin{equation} \label{IC}
p(b') = p_{0 } \,S(b',m)~~~\mbox{with}~~S(b,m)= m b K_1(m\, b );~~~~~
\bar{p}(\vec{b} - \vec{b}) = p_{0} S( \vec{b} - \vec{b}',m) ~~~~~~~z_m = e^{\Delta(1 - p_{0})Y}
\end{equation}

Both $p_{0} $ and mass $m$, as well as the Pomeron intercept
 $\Delta$, are  parameters of the model,  which are determined by 
fitting to the relevant data. Note, that
 $S\Lb b, m\Rb \xrightarrow{m_i\,b \gg 1}\,\exp\Lb - m\,b\Rb$
 in accord with the Froissart theorem\cite{FROI}.

From \eq{IC}  we find that

\bea 
a(b,b')\,\equiv\, a\Lb p, \bar{p}, z_m\Rb &=&\,\h\Lb p + \bar{p}\Rb  \,+\,\frac{1}{2\,z_m}\Lb (1-p)(1-\bar{p} ) \,-\, D\Rb;\label{ALEQ}\\ 
b(b,b') \,\equiv\,b\Lb p, \bar{p}, z_m\Rb \,\,&=&\, \h \frac{p- \bar{p}}{1 - p} -\frac{1}{2 z_m (1 - p)}\Lb (1- p) (1 - \bar{p})  -   D\Rb;\label{BEEQ}\\
~~
D &=& \sqrt{4 p (1 - p) (1-  \bar{p}) z_m -  \Lb (1 - p) (1-  \bar{p}) - (p -  \bar{p}) z_m\Rb^2};\label{D}
\eea
  
  These equation are the explicit solutions to \eq{H03} and \eq{H0BC}.

\par{\it Amplitudes:}
In the following equations $ p \equiv p ( b')$ and $\bar{p}
 \equiv \bar{p}(\vec{b} - \vec{b} ')$.

~
$$z = e^{\Delta\,(1- p_{0})\,y}$$
~

$S (a,b,z) \equiv S(a(b,b'),b(b,b'),z_m)$, \,\, $ X( a, b,z)  \equiv  X(a(b,b'),b(b,b'),z_m)$

\begin{equation}
X(a,b,z)) =  \frac{a + b z}{1 + b z}
\end{equation}

\begin{eqnarray}
&&SS(a,b,z)=\\
&&-(a-1) \text{Li}_2(-b z)+a
   \text{Li}_2\left(-\frac{b
   z}{a}\right)+(a -1)
   \text{Li}_2\left(\frac{a+b
   z}{a -1}\right)+\frac{1}{2} a \log
   ^2((1-a) b z)\nn\\
   && -(a -1) \log (b  z+1)
   \log ((1-a) b z)
  -\left(a \log
   (z)-(a-1) \log \left(-\frac{b
   z+1}{a -1}\right)\right) \log (a + b
   z)\nn\\
   &&  +a \log (z) \log \left(\frac{b
   z}{a+1}\right)\nn
   \end{eqnarray}
   
   \begin{equation} \label{FIN}
   S(a,b,z) \,\,=\,\,SS(a, b, z) \,-\,SS(a , b,z=1)   \end{equation}

~  
     The amplitude is  given by 

   \begin{eqnarray} \label{AIK}
\hspace{-1cm}&&A(s, b)\,\,\,=\,\,\,1\,\,\,-\,\,\,e^{ - \Omega\Lb W, b\Rb             }\,\,=\\
\hspace{-1cm}&&\,1  - \exp\Bigg( \frac{1}{p_{0}}\int \frac{m^2 d^2 b'}{4 \pi} \Big( S(a,b,z_m) \,\,+\,\, a(b,b') \Delta (1 - p_0) Y\Big)  - \int \frac{m^2 d^2 b'}{4 \pi}  \bar{p}( \vec{b} - \vec{b}',m)\,X(a, b,z_m) \Bigg)\nn
 \end{eqnarray}

 %%%%%%%%%%%%%%%%%%%%%%%%%%%%%%%%%%%%%%%%%%%%%%%%%%%
\section{The Odderon contribution}
\subsection{ Odderon exchange}
%%%%%%%%%%%%%%%%%%%%%%%%%%%%%%%%%%%%%%%%%%%%%%%%%%%%%
As has been mentioned, we view the Odderon as a reggeon with negative
 signature and with the intercept
$\alpha_{\rm Odd}(t=0)$=1. Generally speaking its contribution to the
 scattering amplitude has the following form:

\beq \label{ODD1}
O_{i k}(s,b)\,\,=\,\,\eta_{-}(t) \,g^i_{\rm Odd}(b)\,g^k_{\rm Odd}(b)\,\,s^{\alpha_{\rm Odd}(t)\,\,-\,\,1} 
\eeq
where $\eta_{-}$ is a signature factor $\eta\,\,=\,\,\tan\Lb \h \pi\,
\alpha_{\rm Odd}(t)\Rb\,\,-\,\,i$
, $g^i_{\rm Odd}$ is the vertex for the  interaction  of the Odderon with
 state $i$,  and $\alpha_{\rm Odd}$ denotes  the trajectory. The Odderon
  appears naturally in  perturbative QCD. As one can see from
  \fig{odqcd} the QCD Odderon describes the exchange of three
 gluons and all the interactions between them. The QCD Odderon  has
 the trajectory with the intercept equal to 1 and which  does
 not depend on $t$\cite{BLV,KS}. Hence,  the Odderon only contributes
  to the real part of the scattering amplitude. For an 
 estimate we will use the following form of the Odderon contribution:
\beq \label{ODD2}
O_{i k}(s,b)\,\,=\,\,\pm \, \sigma_0 e^{ - \frac{b^2}{4 \,B}}
\eeq
where sign plus corresponds to proton-antiproton scattering, while
 sign minus describes the proton-proton collisions.
The value of $\sigma_0$ was evaluated in Ref.\cite{RYODD} (see also
 Ref.\cite{LERY90}) in the framework of perturbative QCD. It turns
 out that $\sigma_0\,\,=\,\,20.6\,\bas^3\,mb$. In perturbative QCD,
 we expect that $B$ is smaller than for the elastic scattering.  We
 choose $ B=5.6-6 \,GeV^{-2}$  for our estimates\cite{GLPPM,KMRO}.
 In \eq{ODD2} we assume that $g^i_{\rm Odd}(b)$
in \eq{ODD1} does not depend on $i$.
%%%%%%%%%%%%%%%%%%%%%%%%%%%%%%%%%%%%%%%%%%%%%%%%%%%%%
\begin{figure}
\centering
      \includegraphics[width=14cm]{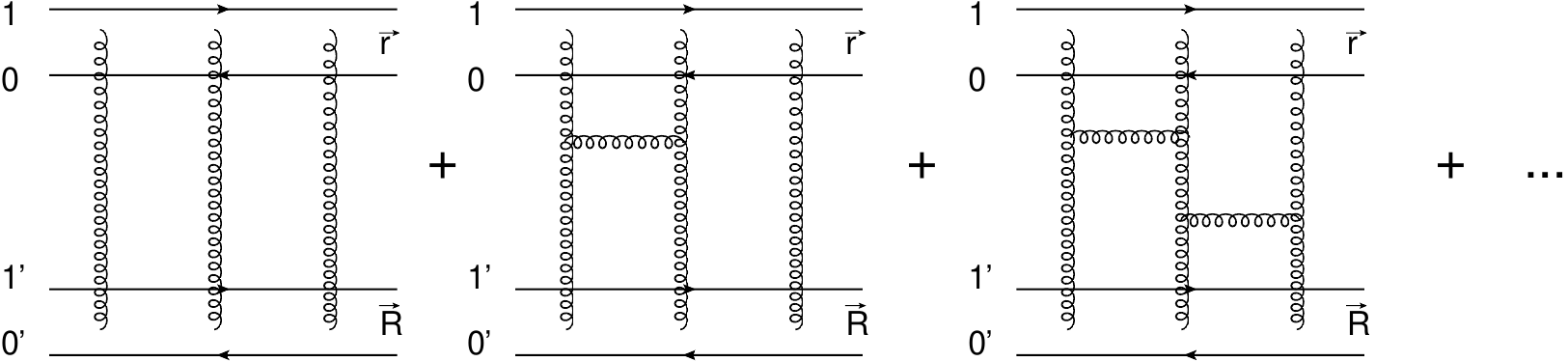} 
         \caption{QCD Odderon for two dipoles scattering: the wavy
 lines describe  gluons and the solid lines correspond to quarks.     }      
\label{odqcd}
\end{figure}
%%%%%%%%%% %%%%%%%%%%%%%%%%%%%%%%%%%%%%%%%%%%%%%%%%%%%%%%%  
   
   %%%%%%%%%% %%%%%%%%%%%%%%%%%%%%%%%%%%%%%%%%%%%%%%%%%%%%%%%  
   \subsection{Shadowing corrections}
   %%%%%%%%%% %%%%%%%%%%%%%%%%%%%%%%%%%%%%%%%%%%%%%%%%%%%%%%%     
     In the eikonal   model the elastic amplitude is equal to
 (see \fig{sc}-a)
   \bea \label{SC1}
   A_{\rm el}\Lb s, b\Rb  \,\,\,&=&\,\,\,\,1\,\,- \,\,\exp\Big(  -\,\,\Omega\Lb s, b\Rb\Big)
   \eea  
   
   \eq{SC1} is the series whose general term is proportional to
 $\Omega^n/n!$.  In the case of  Odderon exchange we
 need to replace one of $\Omega$ by $O(s,b)$. Hence
 $\Omega^n/n!$
  should be replaced by $ O(s,b)  \,n\, \Omega^{n - 1}/n! 
  \,\,=\,\,O(s,b)   \Omega^{n - 1}/(n - 1)!  $. Finally, we
 have (see also Ref.\cite{GLP})
  \beq \label{SC2}
O^{\rm SC}\Lb s, b\Rb  \,\,\,=\,\,\,\, \,O(s,b)\,e^{- \Omega\Lb s, b\Rb} \,\,=\,\,
\,\,O(s,b)\,\Bigg( 1\,\,-\,\,A_{\rm el}\Lb s, b\Rb\Bigg)
 \eeq

   %%%%%%%%%%%%%%%%%%%%%%%%%%%%%%%%%%%%%%%%%%%%%%%%%%%%%
\begin{figure}
\centering
      \includegraphics[width=12cm]{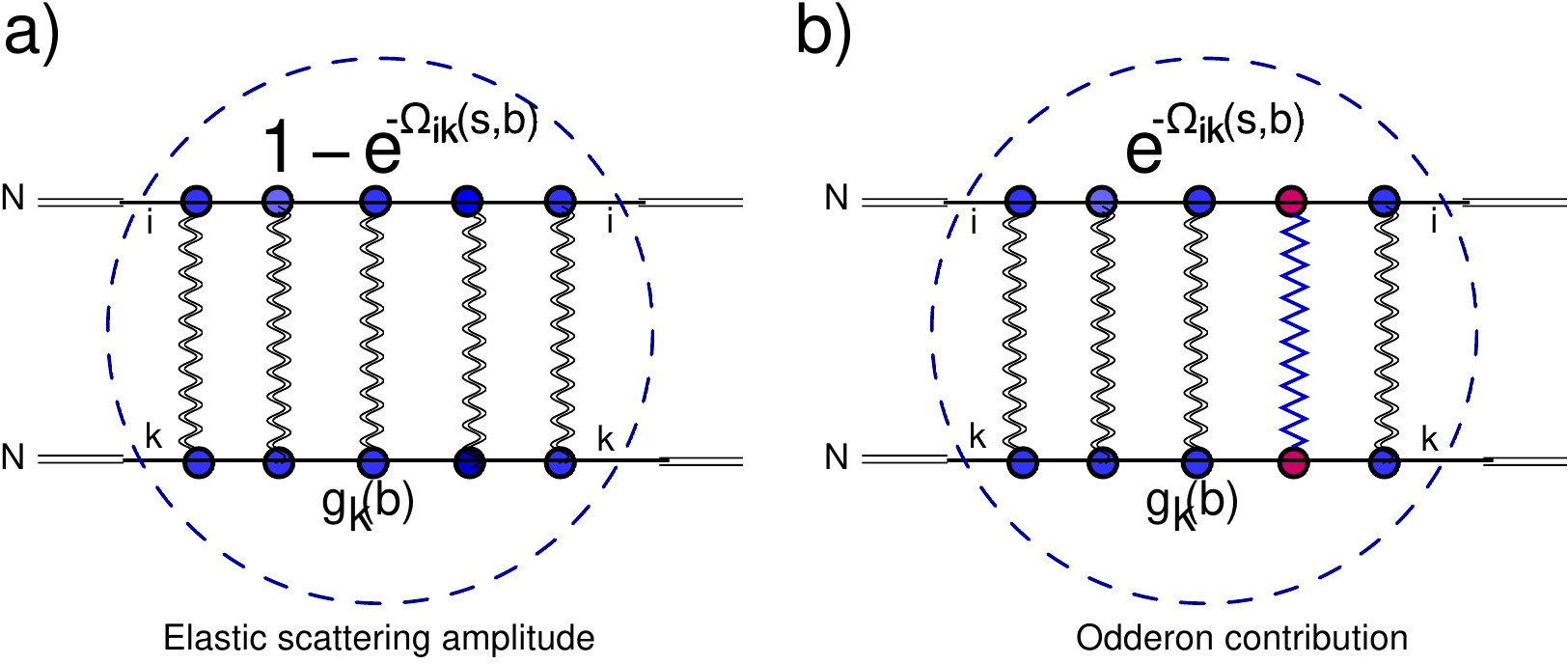} 
         \caption{Shadowing corrections to the Odderon exchange:
 \fig{sc}-a: elastic amplitude in the two channel model.
 \fig{sc}-b: the shadowing corrections in our model.  
 The wavy lines describe the Pomeron exchanges while
 the zigzag line corresponds to the exchange of the Odderon.   }      
\label{sc}
\end{figure}
%%%%%%%%%% %%%%%%%%%%%%%%%%%%%%%%%%%%%%%%%%%%%%%%%%%%%%%%%     
   
      %%%%%%%%%% %%%%%%%%%%%%%%%%%%%%%%%%%%%%%%%%%%%%%%%%%%%%%%%  
   \subsection{Numerical estimates}
   %%%%%%%%%% %%%%%%%%%%%%%%%%%%%%%%%%%%%%%%%%%%%%%%%%%%%%%%%    
   In this section we make estimates using our model
 for $\Omega$, with parameters that are given by Table I.
         %%%%%%%%%%%%%%%%%%%%%%%%%%%%%%%%%%%%%%%%%%%%%%%%%%%%%
\begin{figure}
\centering
      \includegraphics[width=12cm]{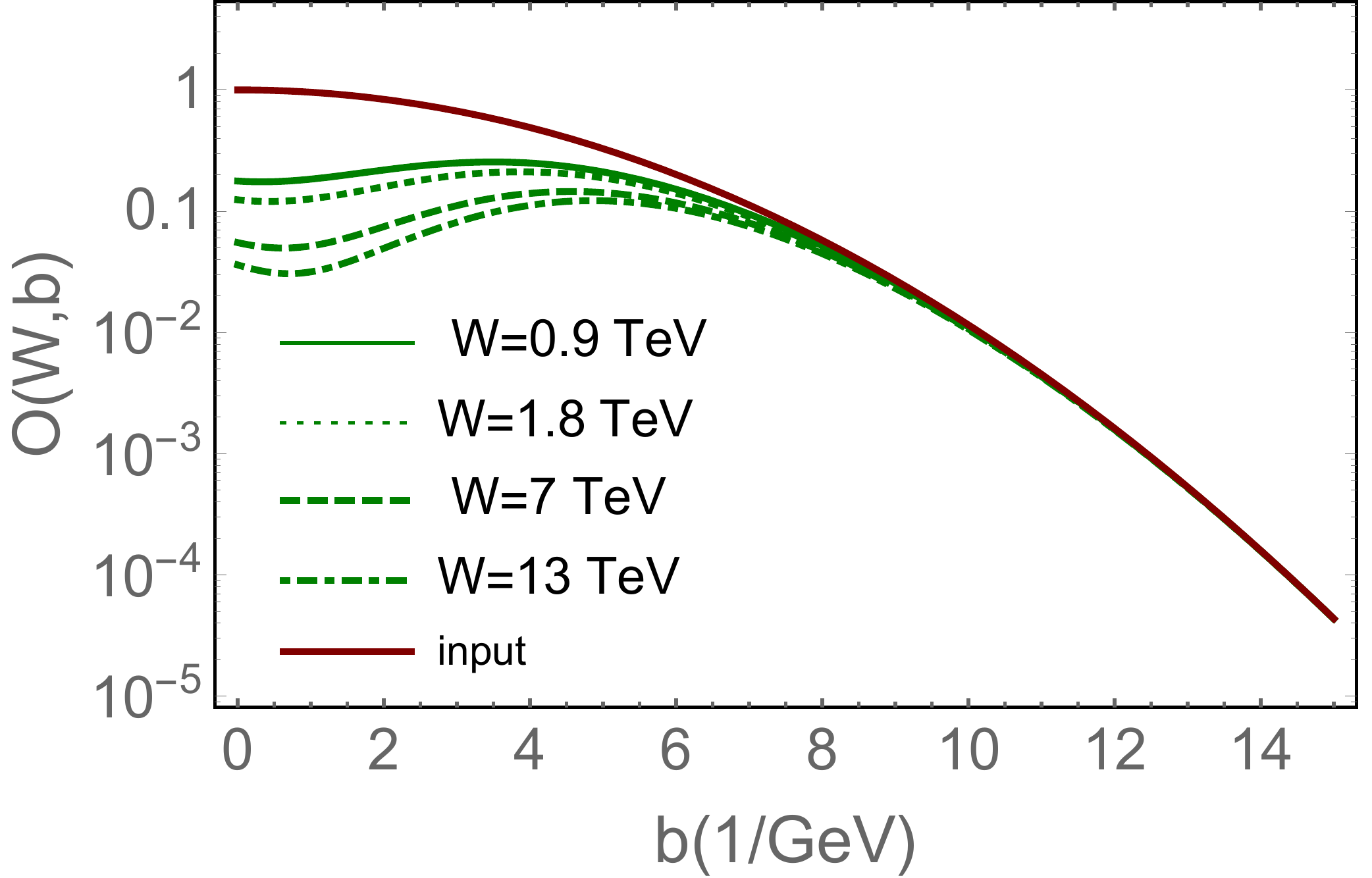} 
         \caption{$O\Lb W, b\Rb$ versus $b$ for different energies.
 The red line corresponds to the contribution of \eq{ODD2}.    }      
\label{b}
\end{figure}
%%%%%%%%%% %%%%%%%%%%%%%%%%%%%%%%%%%%%%%%%%%%%%%%%%%%%%%%%       
     In \fig{b} we plot the $b$ dependence of the Odderon contribution.
 One can see that the shadowing corrections lead to a considerable
 suppression of the Odderon contribution at small $b$  in comparison
 with \eq{ODD2} (see red line in \fig{b}).    This suppression
 is much smaller than in our approach, based on CGC\cite{GLP}. The reason
 for this is  that in our model the value of  $A_{\rm el}\Lb s,
 b\Rb $  
   turns out to be smaller than 1 even at very high energies. Due to this
 $O\Lb W, b=0\Rb\,\,\neq\,\,0$    even at  $W \approx 20 TeV$.

   In \fig{rho} we plot the contribution of the Odderon to the ratio of
 $\rho\,\,=\,\,{\rm Re}/{\rm Im}$ parts of the scattering amplitude as
 function of energy. One sees the influence of the shadowing
 corrections, which induce the energy dependence of this ratio
 on energy. \eq{ODD2} shows that the Odderon   does not depend
 on energy without these corrections.  This induced energy dependence
 turns out to be rather large causing a  decrease of $\rho$
 in the energy range: W = 0.5 $\div$ 20 TeV.  However, this effect is
 much smaller than in our previous estimates \cite{GLP} and the value
 of $\rho$ does not contradict the experimental data
 \cite{TOTEMRHO1,TOTEMRHO2,TOTEMRHO3, TOTEMRHO4}.
    
  The shadowing correction  has a remarkable 
effect on the $t$-dependence
 of the scattering amplitude (see  \fig{q} ). We see that the shadowing 
corrections   lead to a  narrower  distribution over $t$, than the 
input
 given by \eq{ODD2}, which is shown in \fig{q} by the red line.
   
      %%%%%%%%%%%%%%%%%%%%%%%%%%%%%%%%%%%%%%%%%%%%%%%%%%%%%
\begin{figure}
\centering
      \includegraphics[width=12cm]{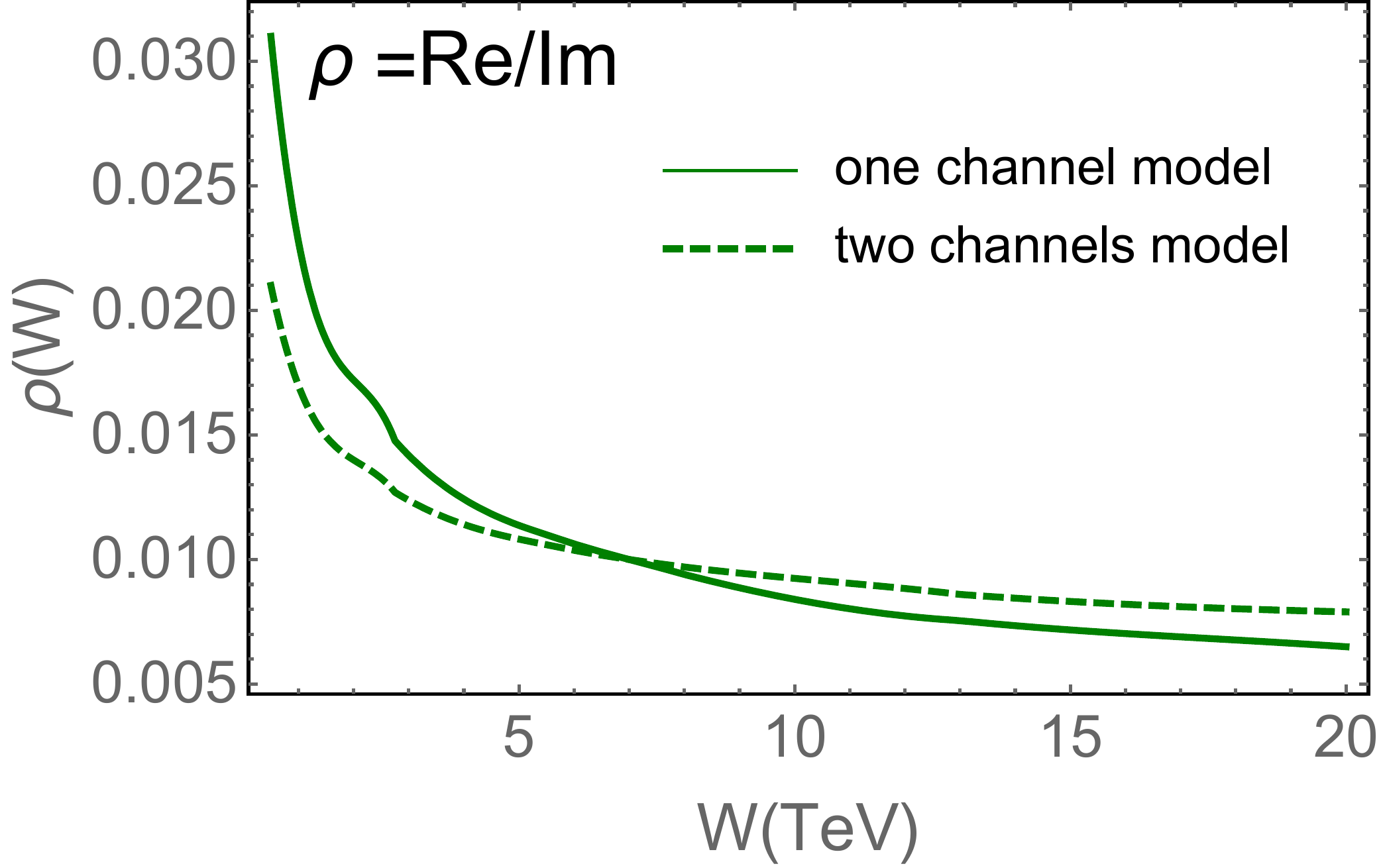} 
         \caption{$\rho = {\rm Re}/{\rm Im}$ due to the Odderon
 contribution versus W in our model. The solid line presents the estimates,
 using \eq{SC2}, while the dashed line describes the evaluation in two 
channel model  of Ref.\cite{GLPPM2}. }      
\label{rho}
\end{figure}
%%%%%%%%%% %%%%%%%%%%%%%%%%%%%%%%%%%%%%%%%%%%%%%%%%%%%%%%%       
   
%%%%%%%%%%%%%%%%%%%%%%%%%%%%%%%%%%%%%%%%%%%%%%%%%%%%%%%%%%%
\begin{table}[h]
\begin{tabular}{|l|l|l|l|}
\hline
\hline
$\Delta_{\rm dressed}$ & $p_{0}$ & $m(GeV)$ & $\chi^2$/d.o.f\\
\hline
0.331 $\pm$ 0.030 & 0.483  $\pm$ 0.030 &0.867 $\pm$ 0.005& 1.3\\
\hline
  \hline
\end{tabular}
\caption{Fitted parameters.$\Delta_{\rm dressed} = \Delta\Lb 1 - p_{0}\Rb$.}

\label{t2}
\end{table}
%%%%%%%%%%%%%%%%%%%%%%%%%%%%%%%%%%%%%%%%%%%%%%%%%%  
      %%%%%%%%%%%%%%%%%%%%%%%%%%%%%%%%%%%%%%%%%%%%%%%%%%%%%
\begin{figure}
\centering
  \includegraphics[width=8cm]{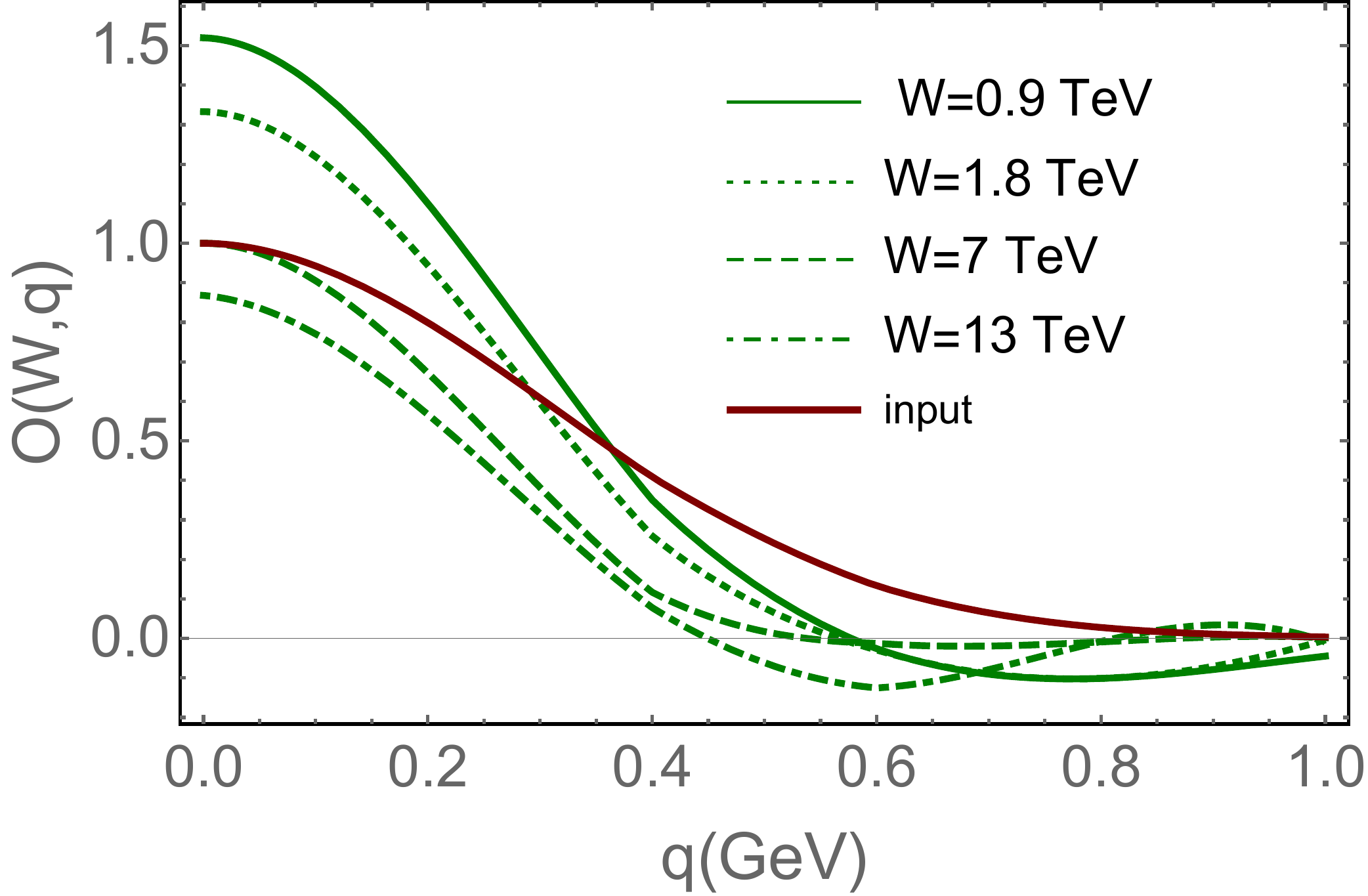}           
               \caption{$O\Lb W,q=\sqrt{|t|}\Rb$ versus $q = 
\sqrt{|t|}$ for different energies. The red line corresponds
 to the contribution of \eq{ODD2}.  }      
\label{q}
\end{figure}
%%%%%%%%%% %%%%%%%%%%%%%%%%%%%%%%%%%%%%%%%%%%%%%%%%%%%%%%% 

\begin{boldmath}
 \section{Dependence of the elastic cross sections on $t$ and the Odderon}
\end{boldmath}
%%%%%%%%%%%%%%%%%%%%%%%%%%%%%%%%%%%%%%%%%%%%%%%%%%%
 In this section we will look at another facet of the Odderon
 contribution: it  could contribute to the real part of the
 scattering amplitude at the value of $t=t_{min}$, where $d 
\sigma_{\rm el}/dt$ has a minimum.

 We  attempt  to describe the elastic cross section for $|t|=0 \div 
1\,GeV^2$ . Our model     predicts the existence of a 
minimum
 in the elastic cross sections, however its position occurs at
 $|t |\,\approx\,0.3\,GeV^2$, which is much smaller than was
 observed experimentally by TOTEM collaboration\cite{TOTEMLT}. 
 
 Assuming that this discrepancy  is due to the simplified 
form of
 $b$ dependence of our amplitude which is given by \eq{IC}, we
 changed the initial conditions of \eq{IC} to the following equations
 \begin{equation} \label{IC1}
p(b') = p_{0 i} \,S(b',m,\mu,\kappa_i)~~~\mbox{with}
~~S(b,m,\mu,\kappa_i)=  \Lb 1 - \kappa\Rb \Lb m\,b\Rb^{\nu_1} K_{\nu_1}(m \,b )\,\,+\,\,\kappa \frac{\Lb m \,b\Rb^{\nu_2} K_{\nu_2}(\mu \,b )}{2^{\nu_2\,-\,1} \,\Gamma\Lb \nu_2\Rb}
\end{equation}

 In  Table II we present the parameters that we  found
 for the fit. \fig{dsdt} shows the comparison with  TOTEM data
 of Ref.\cite{TOTEMLT}. One can see that we  obtain    good agreement
 with the experimental data for $|t|\,<\,|t|_{min}$ and for
 $|t|\,>\, |t|_{min}$. However, for $ |t| \approx\,|t|_{min}$ the real 
part 
of the scattering amplitude  turns out to be small, and we obtain
a value of the $d \sigma_{el}/d \,t$ approximately   an  order of 
 magnitude smaller than the experimental one.  It should be stressed that
 we do not use  any of  the simplified approaches to estimate the 
real part of
 the amplitude, but using our general expression of \eq{AIK}  for $A_{i 
k}\Lb s,t\Rb$, we consider the sum $A_{ik}\Lb s,+ i \epsilon t\Rb +
 A_{ik}\Lb u-i \epsilon,t\Rb$, which corresponds to positive signature,
 and calculated the real part of this sum.

 %%%%%%%%%%%
 In \fig{dsdt} we estimate the contribution of the $\omega$ -reggeon , 
using the 
 description  taken from the paper of Ref.\cite{DOLA}(note the difference
 between green dashed line and the blue solid curve). This
 contribution is  small, and can be neglected.
 %%%%%%%%%%%%%%%%%%%%%%%%%%%%%%%%%%%%%%%%%%%%%%%%%%%
       \begin{figure}[ht]
    \centering
  \leavevmode
      \includegraphics[width=11cm]{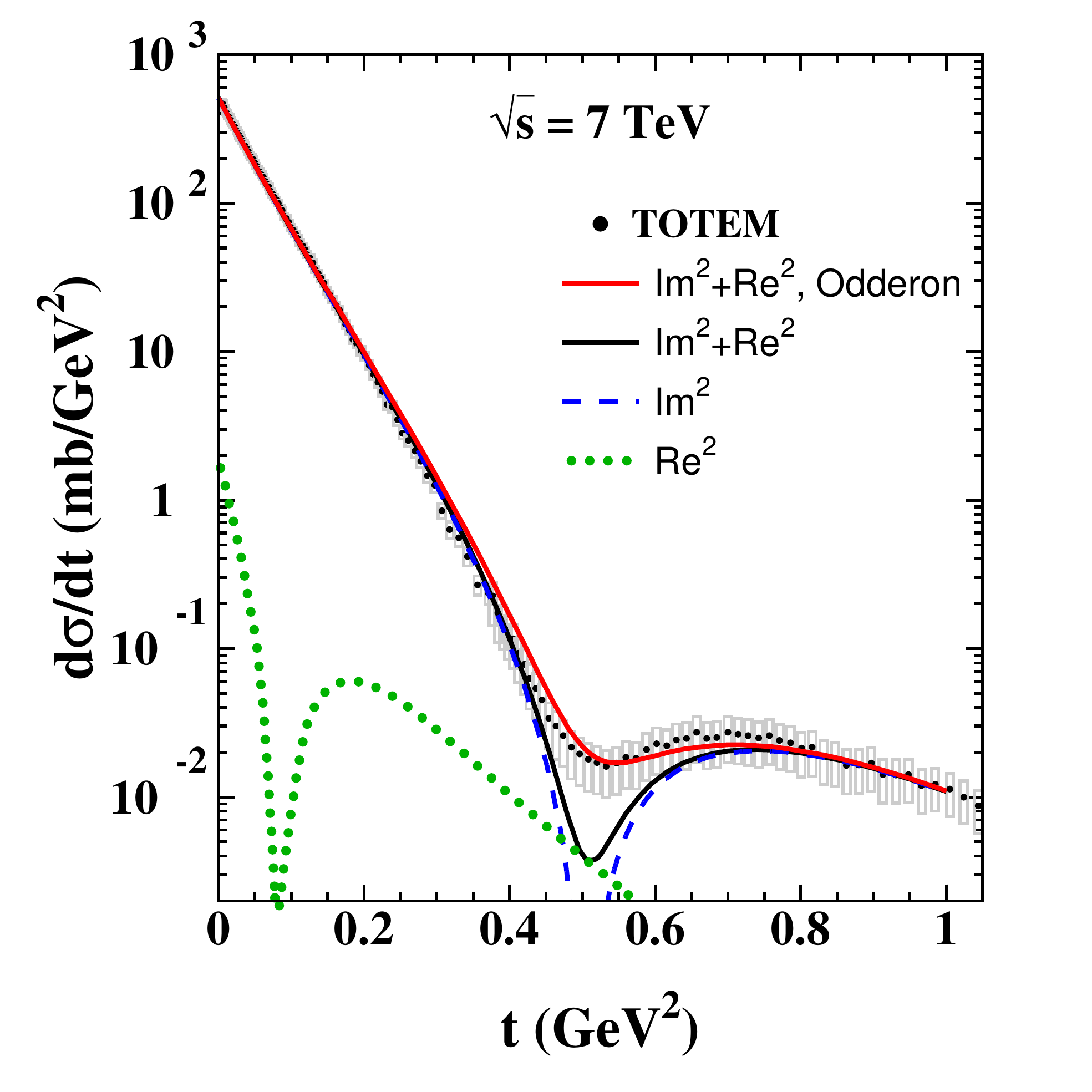}       
             \caption{$d \sigma_{el}/dt$ versus $t$. The black green line
 describes the result of our fit. The dashed  line corresponds to the
 contribution of the imaginary part of the scattering amplitude  to the
 elastic cross section. The dotted line relates to the real part of our
 amplitude.The red solid line takes into account  the contribution of the
 odderon to the real part of the $p p$ amplitude, as is shown in 
 \eq{ODD}.  The data , shown in grey, include systematic errors. They  are
 taken from Ref.\cite{TOTEMLT}. }      \label{dsdt} 
   \end{figure}

 %%%%%%%%%%%%%%%%%%%%%%%%%%%%%%%%%%%%%%%%%%%%%%%%%%%%%%%%%%%%%%%%

To evaluate the real part of the amplitude  we use the relation:  

\beq \label{DSDT1}
{\rm Re}A_{11}(s,t)\,\,=\,\,\h \,\pi\,\frac{\partial}{\partial\,\ln\Lb
 s/s_0\Rb}\, {\rm Im}A_{11}\Lb s,t\Rb|_{\eq{AIK}}
\eeq
\eq{DSDT1} correctly describes the real part of the amplitude only for
 small $\rho={\rm Re}A/{\rm Im} A$. In \fig{dsdt1} we plot
 the $d \sigma/dt$ with  these  estimates for the real part.
  The real part from    \eq{DSDT1} turns
 out to be  almost
 twice  larger than the experimental  data  in the vicinity of $t_{min}$.
  Therefore, at  the minimum, where 
 ${\rm Im}\, A \,\ll\,{\rm Re}A$, \eq{DSDT1} cannot  be used 
 for the real part. However, replacing \eq{DSDT1} by 
\beq \label{DSDT2}
{\rm Re}A_{11}(s,t)\,\,=\,\,\tan\Lb\rho\Rb\, {\rm Im}A_{11}\Lb s,t\Rb|_{ \eq{AIK}}
\eeq
we obtain the same  result, that the real part of the amplitude turns
 out to be too large. Actually,\eq{DSDT2} assumes that the scattering
 amplitude depends on energy as a power $A\Lb s,
 t\Rb\,\propto\,s^{2\,\rho/\pi}$. Our amplitude
 is a rather complex function of energy, and depends
 on $\ln(s)$.
 
   %%%%%%%%%%%%%%%%%%%%%%%%%%%%%%%%%%%%%%%%%%%%
       \begin{figure}[ht]
    \centering
  \leavevmode
      \includegraphics[width=10cm]{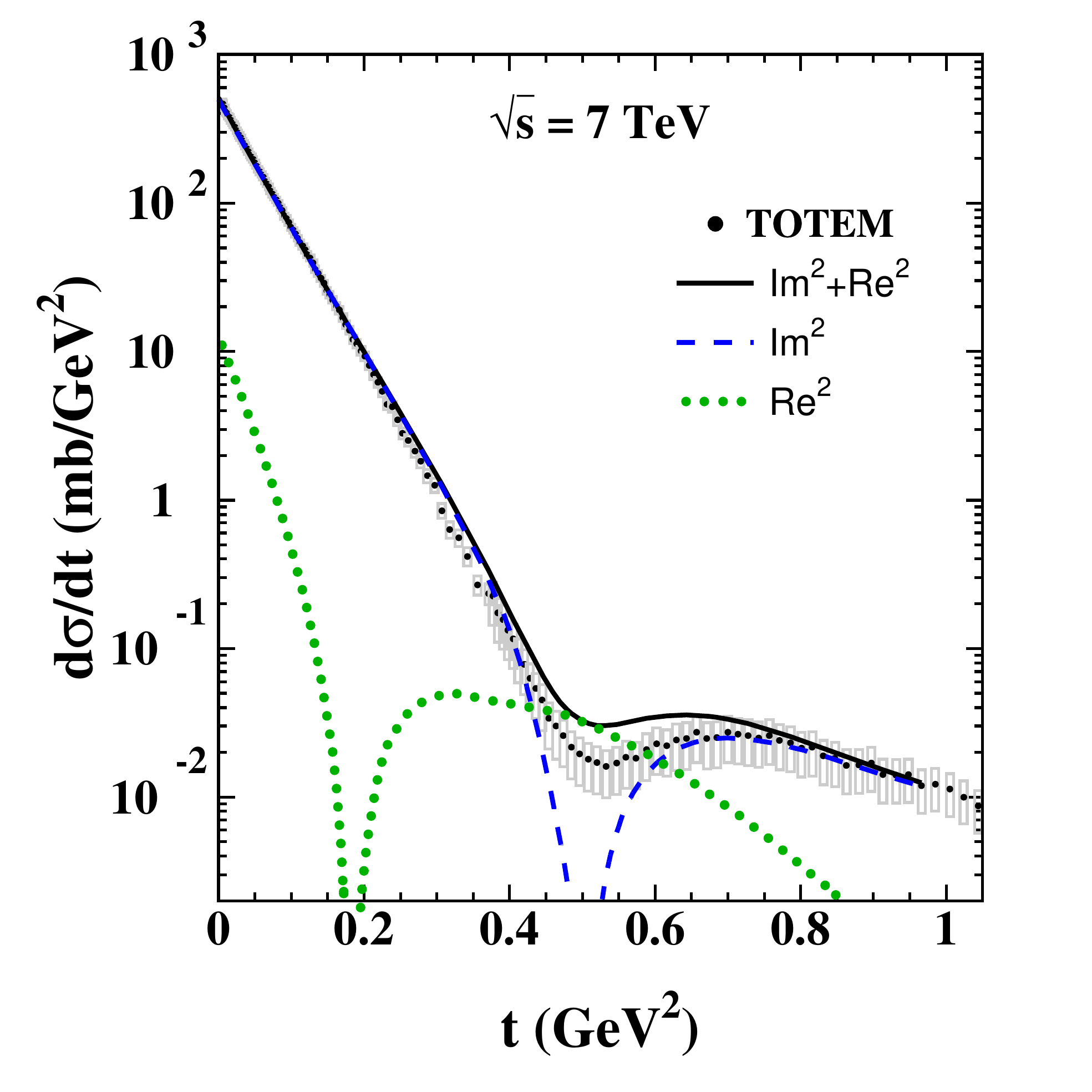}       
             \caption{$d \sigma_{el}/dt$ versus $t$. The solid  line
 describes the result of our fit. The dotted  line corresponds
 to the contribution of the real part of the scattering amplitude
  to the elastic cross section, which is calculated using \eq{DSDT1},
 with added contribution of the
 exchange of the  $\omega$ - reggeon,  which is taken from 
Ref.\cite{DOLA}. We do not show the contribution of the real part
 without the $\omega$-reggeon as  it coincides with the dotted
 line. The dashed line is the contribution of the imaginary part
 of the amplitude.
 The data are taken from Ref.\cite{TOTEMLT}) }      
\label{dsdt1} 
   \end{figure}

 %%%%%%%%%%%%%%%%%%%%%%%%%%%%%%%%%%%%%%%%%%%%%%%%%%%%%%%%%%%%%%%% 

%%%%%%%%%%%%%%%%%%%%%%%%%%%%%%%%%%%%%%%%%%%%%%%%%%%%%%%%%%%
\begin{table}[h]
\begin{tabular}{|l|l|l|l|l|l|l|l|}
\hline
Variant of the fit &$\Delta_{\rm dressed}$ & $p_{0}$   & $m$
 (GeV) &$\mu$(GeV)& $\nu_1$& $\nu_2$&$\kappa$\\
\hline
one channel model  &0.48 $\pm$ 0.01&0.8 $\pm$ 0.05&0.860 &7.6344&0.9&0.1 &0.48\\\hline
\hline
\end{tabular}
\caption{Fitted parameters for $d \sigma_{el}/d t$ 
dependence.$\Delta_{\rm dressed} = \Delta\Lb 1 - p_{01}\Rb$.}
\label{t3}
\end{table}
%%%%%%%%%%%%%%%%%%%%%%%%%%%%%%%%%%%%%%%%%%%%%%%%%%
 Concluding, we see that  to describe the TOTEM experimental data in 
the framework of our model, the contribution to
 the real part of the amplitude from the exchange of the odderon\cite{ODD}
  is needed. 
Hence, our
 estimates confirm the conclusions of Ref.\cite{ODDSC}. In
 \fig{dsdt} we plot the description of the elastic cross section in 
 which we  have added  the odderon contribution to the amplitude of 
\eq{AIK}
 (red solid curve in \fig{dsdt}):
 \beq \label{ODD}
 f\Lb s,t\Rb\,\,=\,\,f\Lb s,t; \eq{AIK}\Rb\,\,\pm\,\, O(s,t) \eeq
 where we consider a QCD odderon\cite{ODD}: the state with odd signature and
  with the intercept  $\alpha_{\rm odd}(t=0)=1$,  which contributes only
 to the real part of the scattering amplitude.  $O(s,t)$ is given by
 \eq{ODD2}. Our odderon parameters are in
 accord with the estimates in Ref.\cite{KMR}. The amplitude $f(s,t)$ is
 related to $a_{el}\Lb s,b\Rb$ as
 \beq \label{OBSEL}
 \frac{d \,\sigma_{el}}{d t}\,\,=\,\,\pi\,|f(s,t)|^2;\,\,\,\,\,a_{el}(s,b)\,\,=\,\,\frac{1}{2 \,\pi}\int d^2 q\, e^{- i \vec{q}\cdot\vec{b}}\,f\Lb s,t\Rb \mbox{where}\,\, t\,=\,- q^2 
 \eeq
 
 In \fig{dsdt2} we show the prediction for proton-antiproton scattering. 
One can conclude that in our model the measurements of the elastic cross
 sections for $p\,p$ and $\bar{p} p$ scattering can provide the estimates 
for
 the odderon contribution. It should be stressed that the contribution
 of the $\omega$-reggeon leads to negligible contribution at
 $W= 7\,TeV$ (see \fig{dsdt1}).

   %%%%%%%%%%%%%%%%%%%%%%%%%%%%%%%%%%%%%%%%%%%%
       \begin{figure}[ht]
    \centering
  \leavevmode
      \includegraphics[width=10cm]{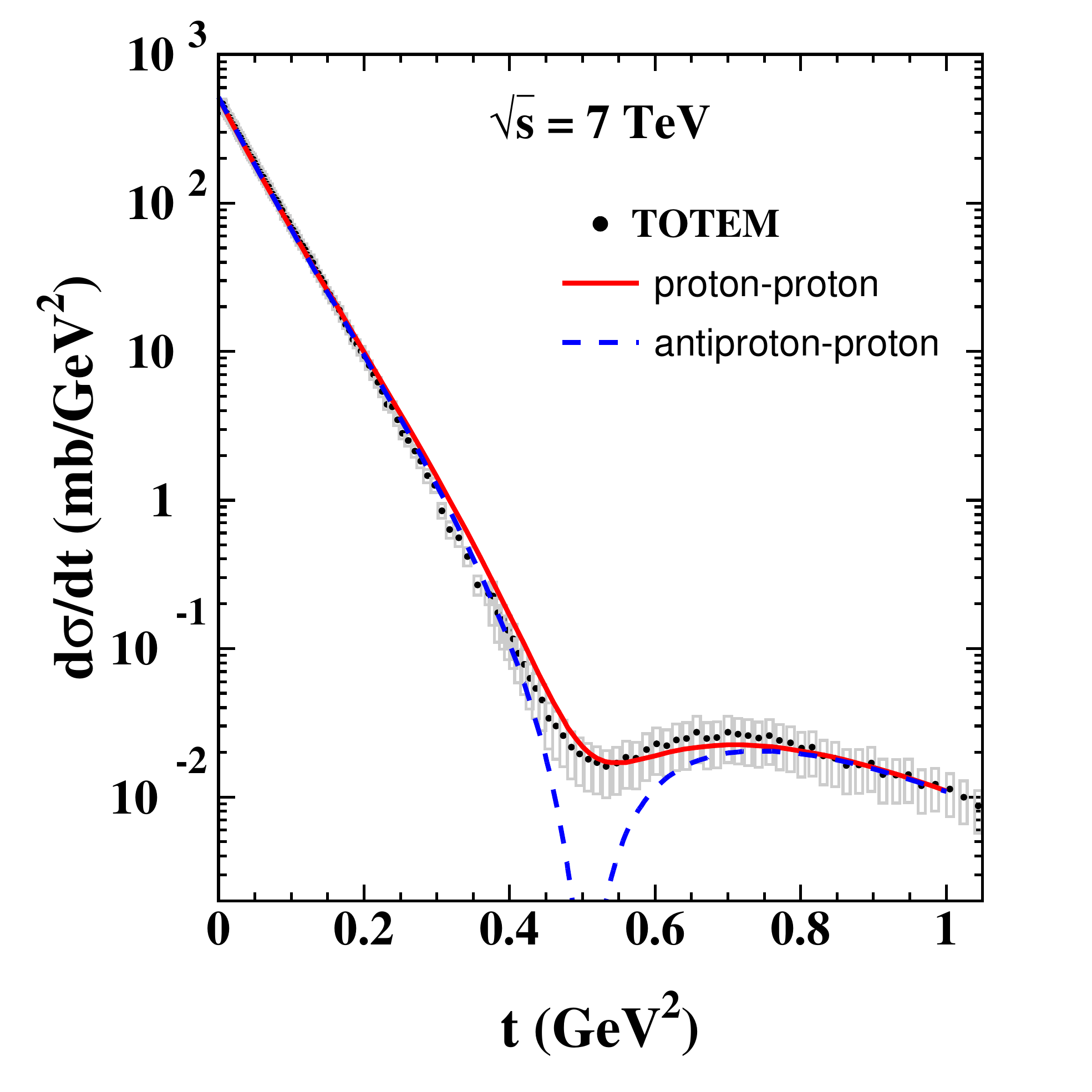}       
             \caption{$d \sigma_{el}/dt$ versus $t$. 
 The solid line describes the elastic cross sections
 for $p p$-scattering  with the odderon contribution
 (see \eq{ODD}), while the dashed line shows the 
elastic cross section for $\bar{p} p$-scattering 
using \eq{ODD}. The data are taken from Ref.\cite{TOTEMLT})
              }      
\label{dsdt2} 
   \end{figure}

 %%%%%%%%%%%%%%%%%%%%%%%%%%%%%%%%%%%%%%%%%%%%%%%%%%%%%%%%%%%%%%%% 

~

~

%%%%%%%%%%%%%%%%%%%%%%%%%%%%%%%%%%%%%%%%%%%%%%%%%%%

 \section{Conclusions}
%%%%%%%%%%%%%%%%%%%%%%%%%%%%%%%%%%%%%%%%%%%%%%%%%%%

In this paper we discussed the Odderon contribution in
 our model\cite{GLPPM} that  provides a fairly good description of 
  $\sigma_{\rm tot}$,$\sigma_{\rm el}$
and $B_{\rm el}$, especially as  related to the energy dependence
 of these observables.  We showed that the shadowing
 corrections are large and induce considerable dependence
 on energy for the Odderon contribution,  which 
 in perturbative QCD  is energy independent . However, this energy
 dependence does not contradict  the experimental data for
 $\rho = {\rm Re}/{\rm Im}$, if we assume that the Odderon
 gives a contribution of about $1 \div 4$ mb at W=7 TeV (see \fig{rhoexp}). 
 
  %%%%%%%%%%%%%%%%%%%%%%%%%%%%%%%%%%%%%%%%%%%%%%%%%%%%%
\begin{figure}
\centering
  \includegraphics[width=8cm]{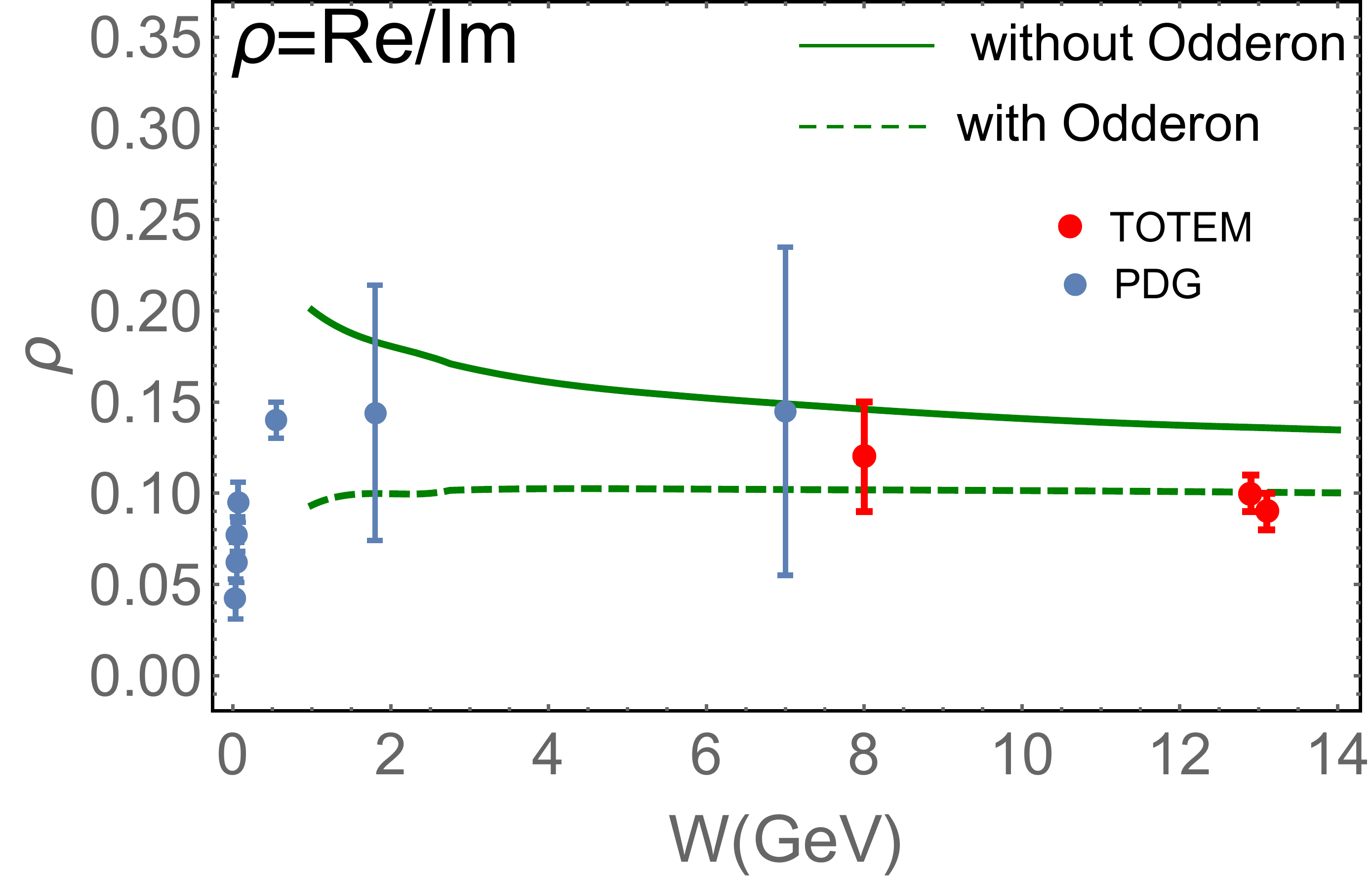}           
               \caption{ $\rho$ = Re/Im for proton-proton scattering
  versus $W =\sqrt{s}$. Data are taken from PDG \cite{PDG} and
 from the TOTEM papers \cite{TOTEMRHO1,TOTEMRHO2,TOTEMRHO3,TOTEMRHO4}. The
 solid line shows the 
 predictions of our model, while the dashed line presents the estimates 
for the
 value of $\rho$, adding the odderon contribution $4 mb$ at W=13 TeV, 
to our model. }      
\label{rhoexp}
\end{figure}
%%%%%%%%%% %%%%%%%%%%%%%%%%%%%%%%%%%%%%%%%%%%%%%%%%%%%%%%%  

 This fact is
 in striking  contrast to  our estimates for the CGC based 
model\cite{GLP}.
 The reason for this difference is that the elastic scattering amplitude
 in our eikonall model does not reach the unitarity limit ($A_{el}\Lb
 W, b=0\Rb=1$, even at very high energies.     The contrast turns out
 to be more pronounced in the two channel model\cite{GLPPM2}, which is shown
 by  dashed lines in \fig{rho}. However, we need to point out here 
that 
the
 comparison with the experimental data in the two channel model is worse
 than in the eikonal one, especially for  $\sigma_{el}$.

 Our attempt to describe the $t$-dependence of the elastic
 cross section shows that we can reproduce the main features of
 the $t$-dependence that are measured experimentally: the slope
 of the elastic cross section at small $t$, the existence of the
 minima in $t$-dependence which is located at $|t|_{min} = 0.52\,GeV^2$
 at W= 7 TeV; and the behaviour of the cross section at $|t|\,>\,|t|_{min}$.
 
 It should be stressed that
 we do not use  any of  the simplified approaches to estimate the 
real part of
 the amplitude which we show ( in our model ), that they do not 
reproduce 
 correctly the  real part of the amplitude at large $t$.
  In our model the real
 part turns out to be much smaller 
 than the experimental one. Consequently,  to achieve a description of the 
data, it is necessary  to add an 
odderon
 contribution.  Hence, our model
 corroborates the conclusion of Ref.\cite{ODDSC}.

 Summarizing, in this paper we have presented estimates
 resulting  from a simple eikonal model,
 which provides a fair description of the data on total and elastic cross 
sections.
 We are aware,  that this is a simplified approach, which  could be a good
 first approximation, but we need to go further. We plan  to
 develop a model which will also describe  diffraction production.
 We have made the first effort \cite{GLPPM2}, but we consider it as
 not very successful, and we need to continue our search. We also
 plan to re-visit our model based on CGC approach \cite{GLP}, with
 the goal to improve it so that , we can also introduce
 the Odderon contribution. We believe that we can base  these attempts
 on  results for QCD Odderon \cite{KOLEB,BLV,KS,LRRW,YHH,CLMS}.

 {\it Acknowledgements.} \\
   We thank our colleagues at Tel Aviv University and UTFSM for
 encouraging discussions.  Our special thanks go to
 Tamas Cs\"org\H o and Jan Kasper for discussion of the odderon contribution
 and elastic scattering   during the Low x'2019 WS. 
 This research was supported  by 
 CONICYT PIA/BASAL FB0821(Chile)  and  Fondecyt (Chile) grants  
1170319 and 1180118 .

 \end{document}